%% file: mcpc.tex
\documentclass[acmsmall]{acmart}

\AtBeginDocument{%
  \providecommand\BibTeX{{%
    \normalfont B\kern-0.5em{\scshape i\kern-0.25em b}\kern-0.8em\TeX}}}

\setcopyright{rightsretained}

\usepackage{graphicx}
\usepackage{appendix}
\usepackage{framed}
\usepackage{booktabs}
\usepackage{tabularx}
\usepackage{CJKutf8} 

\usepackage{enumitem}
\usepackage{pifont}
\usepackage{tcolorbox}
\usepackage{subcaption}
\usepackage{longtable}
\usepackage[ruled,vlined]{algorithm2e}
\usepackage{multirow}
\usepackage[flushleft]{threeparttable} 
\usepackage{tikz}
\usetikzlibrary{trees}
\tikzstyle{every node}=[draw=black,thick,anchor=west]

\usepackage{pstricks}

\usepackage{pgfplots}
\usepackage{pgfplotstable}
\pgfplotsset{compat=1.14}

\usepackage{enumitem}

\usepackage{booktabs}

\usepackage{msc}
\drawframe{no}

\usepackage{wrapfig,lipsum,booktabs}
\usepackage{listings}

\definecolor{codegreen}{rgb}{0,0.6,0}
\definecolor{codegray}{rgb}{0.5,0.5,0.5}
\definecolor{codepurple}{rgb}{0.58,0,0.82}
\definecolor{backcolour}{rgb}{1,1,1}
\usepackage[all]{nowidow}

\usepackage{colortbl}
\usepackage{arydshln}
\usepackage{calculator}
\usepackage{fp}
\usepackage{xcolor}
\usepackage{pgf}
\usepackage{collcell}
\usepackage[export]{adjustbox}
\usepackage{multirow} 
\usepackage{pdflscape} 
\usepackage{booktabs}
\usepackage{colortbl}
\definecolor{SkyBlue}{rgb}{0.53, 0.81, 0.92}
\usepackage{tcolorbox}
 \usepackage{tcolorbox}
\usepackage{mdframed}

\usepackage[T1]{fontenc}

\usepackage{icomma}

\newcommand*{\ApplyGradient}[3]{%
    \pgfmathsetmacro{\PercentColor}{100.0*(#3/(#2*1.0))/#1}%
    \edef\x{\noexpand\cellcolor{SkyBlue!\PercentColor}}\x\textcolor{black}{#3}%
}%
\newcommand{\subHeatMapA}[1]{\ApplyGradient{100}{85}{#1}}
\newcolumntype{A}{>{\collectcell\subHeatMapA}{r}<{\endcollectcell}}

\newcommand{\subHeatMapP}[1]{\ApplyGradient{100}{1}{#1}}
\newcolumntype{P}{>{\collectcell\subHeatMapP}{r}<{\endcollectcell}}

\newcommand{\subHeatMapD}[1]{\ApplyGradient{100}{45}{#1}}
\newcolumntype{D}{>{\collectcell\subHeatMapD}{r}<{\endcollectcell}}

\newcommand{\ignore}[1]{}

\newcommand{\mcpc}{\textsf{MCPCrawler}}

\usepackage{hyperref}

\hypersetup{
    colorlinks = true,
    linkcolor = blue,
    anchorcolor = blue,
    citecolor = blue,
    filecolor = blue,
    urlcolor = blue
}

\begin{document}
\title{A Measurement Study of Model Context Protocol Ecosystem}

\author{Hechuan Guo}
\affiliation{%
  \institution{Shandong University}
  \country{China}
}
\email{hcguo@sdu.edu.cn}

\author{Yongle Hao}
\affiliation{%
  \institution{Shandong University}
  \country{China}
}
\email{ylhao@mail.sdu.edu.cn}

\author{Yue Zhang}
\affiliation{%
  \institution{Shandong University}
  \country{China}
}
\email{zyueinfosec@sdu.edu.cn}

\author{Minghui Xu}
\affiliation{%
  \institution{Shandong University}
  \country{China}
}
\email{mhxu@sdu.edu.cn}

\author{Peizhuo Lv}
\affiliation{%
  \institution{Nanyang Technological University}
  \country{Singapore}
}
\email{peizhuo.lyu@ntu.edu.sg}

\author{Jiezhi Chen}
\affiliation{%
  \institution{Shandong University}
  \country{China}
}
\email{chen.jiezhi@sdu.edu.cn}

\author{Xiuzhen Cheng}
\affiliation{%
  \institution{Shandong University}
  \country{China}
}
\email{xzcheng@sdu.edu.cn}



 \renewcommand{\shortauthors}{Author Name}
 
\begin{abstract}
The Model Context Protocol (MCP) has been proposed as a unifying standard for connecting large language models (LLMs) with external tools and resources, promising the same role for AI integration that HTTP and USB played for the Web and peripherals. Yet, despite rapid adoption and hype, its trajectory remains uncertain. Are MCP marketplaces truly growing, or merely inflated by placeholders and abandoned prototypes? Are servers secure and privacy-preserving, or do they expose users to systemic risks? And do clients converge on standardized protocols, or remain fragmented across competing designs?
In this paper, we present the first large-scale empirical study of the MCP ecosystem. We design and implement MCPCrawler, a systematic measurement framework that collects and normalizes data from six major markets. Over a 14-day campaign, MCPCrawler aggregated 17,630 raw entries, of which 8,401 valid projects (8,060 servers and 341 clients) were analyzed. Our results reveal that more than half of listed projects are invalid or low-value, that servers face structural risks including dependency monocultures and uneven maintenance, and that clients exhibit a transitional phase in protocol and connection patterns. Together, these findings provide the first evidence-based view of the MCP ecosystem, its risks, and its future trajectory.
\end{abstract}

\maketitle

\input{sections/sec01-introduction}

\input{sections/sec02-background}

\input{sections/sec03-research-questions.tex}

\input{sections/sec04-architecture}

\input{sections/sec04-performance}

\input{sections/sec05-market}

\input{sections/sec06-server} 

\input{sections/sec07-client}

\input{sections/sec08-discussion}

\input{sections/sec09-related}

\input{sections/sec10-conclusion}
 
\begingroup
\raggedright
\bibliography{mcpc} 
{\small
\bibliographystyle{IEEEtran}
}
\endgroup
\end{document}

%% file: sections/sec01-introduction.tex
\section{Introduction}
\label{section:introduction}

The recent surge of large language model (LLM)–powered applications~\cite{openai2023plugins,anthropic2024claude} has brought new demands for interoperability, modularity, and extensibility. The Model Context Protocol (MCP) has emerged as a promising standard to meet these needs, offering a lightweight interface for connecting clients, servers, and external resources~ \cite{introducingmcp,li2025largelanguagemodelstrusted,li2025urgentlyneedprivilegemanagement}. MCP specifies a lightweight yet expressive mechanism for describing, discovering, and invoking external capabilities within a model’s operational context. Its ambition is to become for LLM integration what HTTP was for the Web \cite{fielding2000architectural} or what USB became for peripheral devices \cite{windowsplugplay}, a unifying protocol layer that ensures portability and reusability across diverse platforms. The significance of such a standard has also been recognized in broader AI governance contexts: international standardization bodies such as ISO/IEC JTC 1/SC 42 have emphasized the need for interoperability and sustainable integration frameworks for AI technologies \cite{isoiec2025sc42}.\looseness=-1

However, despite its promise, the trajectory of the MCP ecosystem remains highly uncertain. Marketplaces may list thousands of entries, but it is unclear how many represent actively maintained, production-quality projects. For instance, users on Reddit have questioned whether MCP is simply another overhyped standard destined to fade once the novelty wears off, noting that ``\textit{most of MCPs, like sequential thinking, don’t really need to be MCP and are not a good fit}''~\cite{reddit2025mcp}. Without a systematic perspective, it is difficult to determine whether MCP is undergoing genuine, sustainable growth or merely experiencing a transient surge driven by novelty and hype.
At the same time, MCP servers frequently integrate with sensitive resources such as authentication systems, proprietary APIs, and personal data connectors. While such deployments could enable secure interoperability, they also risk exposing users to privacy and security vulnerabilities if not carefully safeguarded. MCP clients, which act as the bridge between servers and LLMs, further influence the ecosystem’s trajectory through their choice of communication protocols and connection modes. Some developers advocate for SSE as the natural default for future interoperability, while others favor the simplicity of stdio for debugging and lightweight use.

Taken together, these uncertainties highlight the need for a structured, measurement-based study of the MCP ecosystem. In this work, we provide the first systematic analysis of MCP markets, servers, and clients. To guide our investigation, we focus on three research questions:

\begin{itemize}
    \item \textbf{RQ1 (Market):}  What is the current scale of the MCP ecosystem across major markets, and what do observed growth patterns suggest about its future trajectory?

 \item \textbf{RQ2 (MCP Server):} To what extent is the MCP ecosystem secure and privacy-preserving, and how do structural factors, and functional roles collectively shape its overall risk posture?

 \item 
\textbf{RQ3 (MCP Client):} How do the interaction protocols and connection modes of MCP clients shape the evolutionary trajectory of the ecosystem?
\end{itemize}

To answer these research questions, we designed and implemented MCPCrawler, the first systematic measurement framework for the MCP ecosystem. MCPCrawler was built to collect, normalize, and analyze data from multiple heterogeneous markets that each publish their own listings of MCP servers and clients. Specifically, it works in three stages. First, it discovers and aggregates entries from marketplaces, applying rule-based noise filtering to exclude invalid or low-value records such as inactive forks, placeholder repositories, or projects without executable code. Second, it extracts metadata from valid entries, including declared dependencies, repository activity, implementation language, functional category, communication protocol, and connection patterns. Finally, it normalizes and visualizes this information across markets, enabling comparative measurement of ecosystem scale, security posture, and interoperability dynamics.

Over a 14-day campaign, MCPCrawler collected 8,401 distinct entries, including 8,060 valid MCP servers and 341 valid MCP clients. This dataset allows us to study the MCP ecosystem across three complementary dimensions  (markets, servers, and clients) and to provide the first evidence-based view of its current state, which directly address the three research questions.

\begin{itemize}
\item (\textbf{Market -- Ecosystem Scale and Growth Potential).}  The MCP ecosystem is sizable but fragile. Across six major markets, \mcpc{} collected 17,630 raw entries, but after filtering, only 8,656 (49.1\%) were valid.  For example, in MCP.so, just 7,223 of 16,646 (43.4\%) server records passed validation, while MCP Market fared even worse at 26.4\% (3,765 of 14,280).  Longitudinal analysis shows that MCP.so has largely plateaued, whereas MCP Market contributes most of the ongoing growth. 
Yet, more than 50\% of the ecosystem consists of placeholders, forks, or abandoned projects,  raising doubts about sustainability. Moreover, overlap analysis reveals both redundancy and fragmentation: while 41.9\% of projects appear in multiple markets, only 6.9\% are indexed in four or more, showing that no single market provides full coverage.

\item \textbf{(Server -- Security and Privacy Posture).}  
Our analysis of 8,060 valid MCP servers highlights multiple structural risks. On dependencies, we found strong monocultures: e.g., Java servers overwhelmingly use Spring (spring-boot, spring-core, spring-web), meaning vulnerabilities such as SpringShell could cascade widely. Python and TypeScript servers rely on \texttt{pydantic} and \texttt{zod}, which improve input safety, but Go and Rust servers lack equivalent safeguards, relying on manual validation. Maintenance also varies: 40.9\% of servers were updated within 90 days, but 21.9\% had been inactive for over a year, creating a long tail of unpatched projects. 
Functionally, 11.2\% of servers contain code that potentially invokes sensitive APIs, with authentication-related services comprising 43\% of this group. Their inclusion of high-impact APIs increases the potential consequences of misconfiguration or compromise.
Together, these findings show that while good practices exist, the ecosystem remains highly vulnerable to 
supply-chain attacks, abandonment risks, and privacy exposures.

\item \textbf{(Client -- Connection Patterns in Ecosystem Evolution).}  
From 341 valid MCP clients, we find evidence of both convergence and fragmentation. 
On protocols, SSE dominates with 56.9\% (194 clients), followed by stdio at 38.1\% (130), 
while others remain marginal (4.9\%). 
This indicates a shift toward SSE as a de facto standard, yet the persistence of stdio suggests 
that diverse design philosophies remain relevant, especially for lightweight or local scenarios. 
On connection modes, 80.9\% of clients support only a single server connection, 
while 19.1\% (65 clients) allow multiple concurrent connections. 
This skew shows that most clients still favor point-to-point simplicity, 
but a significant minority are evolving toward multi-server integration, 
enabling richer workflows and redundancy. These patterns suggest the ecosystem is at a transitional phase, with SSE and single-connection models dominating today, 
but multi-connection and protocol diversity hinting at possible future directions.

\end{itemize}

\paragraph{Contribution.}
In summary, this paper makes the following contributions:

\begin{itemize}
    \item \textbf{First large-scale empirical study.}  
    We present the first large-scale empirical study of the MCP ecosystem, covering six major markets and systematically analyzing 8,060 servers and 341 clients.
    
    \item \textbf{Measurement framework and dataset.}  
    We design and implement \mcpc{}, a dedicated measurement framework for discovering, filtering, and analyzing MCP projects, and release the resulting dataset to support reproducible research and future studies. 
    
    \item \textbf{New empirical findings.}  
    We uncover several new findings about the MCP ecosystem: markets are fragmented and inflated by low-value entries; servers exhibit uneven maintenance, supply-chain monocultures, and exposure of sensitive APIs; and clients show a transitional pattern, with SSE emerging as a dominant protocol but diversity persisting in connection modes. These findings provide the first evidence-based insights into the scale, risks, and evolutionary trajectory of MCP.
\end{itemize}

\noindent\textbf{Dataset and Tool.} \mcpc{} is open-sourced at \url{https://github.com/zhuaiballl/mcpc}.  The collected dataset has been made publicly available at \url{https://github.com/zhuaiballl/mcp\_collection}.

%% file: sections/sec02-background.tex
\section{Background} 
\label{section:background}

\subsection{MCP Overview}
\label{subsec:architecture}

The Model Context Protocol (MCP) is an emerging open standard for connecting large language model (LLM) applications to external data sources, tools, and services \cite{anthropic2024mcp,tajwar2024preferencefinetuningllmsleverage,openaitoolcalling,surveymcp}. 
It serves as a ``standardization layer'' that lets an LLM interact with live data and APIs much like a USB-C port links a computer to peripherals \cite{wei2022emergent,cai2024llmtool}. 
MCP treats the AI application as an MCP Host that can ``mount'' external MCP servers via one or more MCP clients \cite{glama2024doc,smithery2024market}. 
Each MCP server exposes contextual resources (data), prompts (templates), and tools (functions) to the connected AI system, enabling the model to retrieve fresh information and invoke actions securely \cite{qin2024survey,iqbal2024llmplatformsecurityapplying,bommasani2021opportunities}. 
For example, an MCP server might expose a ``weather/current'' tool or a ``database/schema'' resource; the AI application's MCP client discovers these offerings and incorporates them into the model's toolset \cite{yao2023react,schick2023toolformer}. \looseness=-1

\begin{wrapfigure}[13]{r}{0.6\columnwidth}
\centering

    \includegraphics[width=1\linewidth]{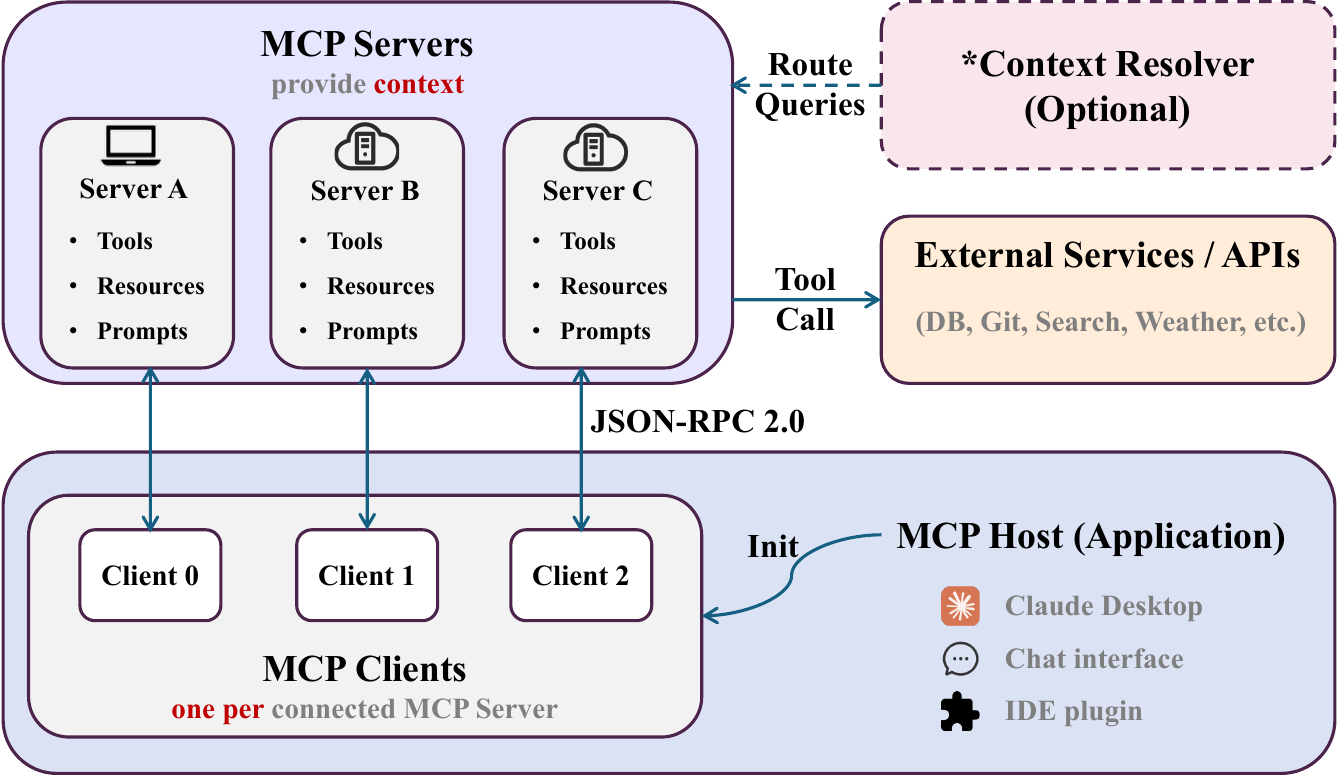}
    \caption{A typical MCP architecture}
    \label{fig:mcp_architecture}
\end{wrapfigure}

\vspace{2mm}
\noindent\textbf{MCP Architecture.} 
MCP is designed around a client--server paradigm that leverages stateful JSON-RPC 2.0 communication \cite{jsonrpc,fielding2000architectural}.
Rather than being a monolithic framework, MCP provides a structured way for LLM-equipped applications to discover, access, and coordinate external resources. This design reflects the broader trend of treating LLMs not only as text generators but also as orchestrators that interact with heterogeneous tools and services. 
As shown in \autoref{fig:mcp_architecture}, key components of the architecture include: 
\begin{itemize}
    \item \textbf{MCP Host (Application):} The LLM-equipped application (e.g., Claude Desktop, a chat interface, or an IDE plugin) that initiates connections to servers. The host creates a separate MCP client for each server it connects to \cite{anthropic2024mcp,glama2024doc}.
    \item \textbf{MCP Client:} A client object embedded in the host, responsible for a one-to-one connection with an MCP server. A host connecting to $n$ servers will have $n$ MCP client instances \cite{openaitoolcalling}.
    \item \textbf{MCP Server:} A service (local or remote) that provides contextual data and executable tools \cite{iqbal2024llmplatformsecurityapplying,tajwar2024preferencefinetuningllmsleverage}. Servers can run locally using stdio transport or on cloud infrastructure via streamable HTTP \cite{jsonrpc}.
    \item \textbf{(Optional) Context Resolver:} Some advanced agent frameworks introduce a higher-level resolver that routes queries to the most relevant server based on the domain of the request \cite{liu2023agentbench,guo2024llmbased}. For example, if a user asks “summarize the latest customer support tickets”, the resolver can direct the request to a server connected to the company’s ticketing database (e.g., Zendesk API). Conversely, a request such as ``\textit{generate a UML diagram from this code snippet}'' would be dispatched to a code-analysis server or IDE plugin. By abstracting this routing logic, the resolver allows the host to handle heterogeneous queries seamlessly without hardcoding which server should be called. \looseness=-1
\end{itemize}


\noindent\textbf{MCP Workflow.} The MCP workflow captures the end-to-end interaction pattern by which an LLM, via its client, establishes connections, discovers tools, and invokes server functions to fulfill user requests.
The LLM, acting on the user’s natural language instruction, delegates execution to the MCP client, which initializes the connection, exchanges protocol versions and capability sets with the server, and subscribes to notifications \cite{glama2024doc,anthropic2024mcp}.
After initialization, the client performs tool and resource discovery (e.g., via $\mathsf{tools/list}$ or $\mathsf{resources/list}$ requests) \cite{iqbal2024llmplatformsecurityapplying,yao2023react}.
The host application aggregates these listings into a unified tool registry that the LLM can query when deciding how to fulfill user requests.
When the user issues a task (e.g., “find recent papers” or “summarize a document”), the LLM selects a tool and triggers its execution by sending a $\mathsf{tools/call}$ request with the tool’s URI and arguments. The server executes the corresponding function (such as querying a database or performing a web search) and returns a structured response \cite{schick2023toolformer,cai2024llmtool}, which is then processed by the LLM and surfaced back to the user in natural language.
In addition, servers can proactively push real-time notifications when their available primitives change, ensuring that both the LLM and user always operate over the latest tool state \cite{jsonrpc}.

\vspace{2mm}
\noindent\textbf{Deployment and Implementation of MCP. } 
MCP relies on two transport layers to accommodate different deployment settings: for local servers, stdio offers low-latency communication without network overhead, while for remote servers, HTTP/S with optional Server-Sent Events supports streaming and standard authentication \cite{anthropic2024mcp,glama2024doc}. Beyond transport, MCP is inherently stateful: once initialized, the client and server maintain a session that allows multiple RPC calls to be issued over the same channel \cite{tajwar2024preferencefinetuningllmsleverage}. This design enables AI applications to ``carry their context'' across tools and environments, ensuring interoperability in heterogeneous ecosystems \cite{cai2024llmtool,hassouna2024llmagentumf}. Building on this foundation, different implementations have emerged that extend MCP’s core. Some projects enrich the protocol into agent frameworks with advanced features such as namespace resolution or decentralized data networks \cite{liu2023agentbench,guo2024llmbased}, while others embed reasoning logic, delegating sub-questions back to the model for refinement \cite{yao2023react,schick2023toolformer}. Despite such variations, all implementations adhere to the same underlying handshake, discovery, execution, and notification flows that define MCP.



\looseness=-1

\looseness=-1

\subsection{MCP Markets}
\label{sub:markets}

\begin{wraptable}[9]{r}{0.75\columnwidth}
    \centering
    \caption{Comparison of MCP Marketplaces (data collected in 2025).}
    \label{tab:mcp_marketplaces}
    \scriptsize
    \setlength{\tabcolsep}{3pt}
    \renewcommand{\arraystretch}{0.9}
    \begin{tabular}{l l l l r}
        \toprule
        \textbf{Marketplace} & \textbf{API/SDK } & \textbf{Deployment} & \textbf{Versions} & \textbf{\# Servers} \\
        \midrule
        Smithery         & CLI installer, TypeScript SDK & Local + hosted    & Latest spec           & 5,625 \\
        MCP.so           & Web UI (no public API)        & Web browser       & Multi-version         & 15,704 \\
        Glama            & Web UI + REST API             & Hosted (cloud)    & Latest spec           & 7,675 \\
        MCP Market       & Web UI + JSON API             & Web browser       & Official + community  & 13,830 \\
        Cursor Directory & Web UI                        & Web + IDE         & Latest spec           & 1,560 \\
        PulseMCP         & Web UI (newsletter)           & Web browser       & Community entries     & 5,264 \\
        \bottomrule
    \end{tabular}
\end{wraptable}

A central part of the MCP ecosystem is the emergence of dedicated marketplaces (sometimes referred to as registries), which act as catalogs for MCP-compatible servers and clients. These platforms provide indexing, categorization, and in some cases direct hosting of connectors \cite{smithery2024market,glama2024doc,pulsemcp2024}. By doing so, they reduce duplication of effort and enable developers or agent frameworks to reuse existing connectors rather than building new ones from scratch—mirroring the dynamics seen in plugin ecosystems for web browsers, IDEs, and LLM agents \cite{liu2023agentbench,li2024survey}. Well-known examples include Smithery, MCP.so, Glama.ai, PulseMCP, MCP Market, and Cursor Directory \cite{smithery2024market,glama2024doc,pulsemcp2024}.



MCP marketplaces adopt different deployment models, reflecting a trade-off between local control and cloud-managed convenience. Some markets support \textit{local mode}, where server code is fetched and run in the user’s own environment, giving users full control of credentials and runtime \cite{smithery2024market,glama2024doc}. 
Others provide \textit{hosted mode}, offering cloud-hosted MCP servers with managed runtime and stable endpoints, which reduce setup cost but delegate execution control to the hosting provider \cite{glama2024doc,pulsemcp2024}. 
This design resembles trends in serverless computing and API marketplaces \cite{buyya2008market,javed201652}. 
Other registries, such as MCP.so or Cursor Directory, primarily focus on discovery and indexing rather than deployment \cite{pulsemcp2024}.  
\autoref{tab:mcp_marketplaces} summarizes the major MCP marketplaces, comparing their API/SDK support, deployment options, protocol coverage, and scale of discoverability \cite{smithery2024market,glama2024doc}.
These marketplaces collectively illustrate how MCP is evolving into a multi-platform ecosystem similar to established software markets \cite{buyya2008market,li2024survey}.

		
		


		

%% file: sections/sec03-research-questions.tex
\section{Motivation and Research Questions}
\label{sub:research_questions}

\subsection{Motivation}

As discussed in \S\ref{section:background}, MCP ecosystem has rapidly expanded with the rise of LLM-powered applications and tool integration frameworks. As marketplaces for MCP servers and clients continue to emerge, they create a rich ecosystem of connectors that support diverse functionality, from data retrieval and automation to software development workflows. 

However, this rapid growth also raises several open challenges: First, despite the hype surrounding MCP, it remains unclear whether the ecosystem is experiencing genuine, sustainable growth or is still in an early, experimental stage. Many developers and users appear to adopt a wait-and-see attitude: marketplaces advertise thousands of entries, yet it is not obvious how many represent active, well-maintained projects versus placeholders, experiments, or abandoned efforts.   Second, MCP servers often interface with sensitive resources, such as authentication services, personal information, or proprietary model APIs. Without systematic scrutiny, it is difficult to assess whether existing deployments adopt robust safeguards or inadvertently expose users to privacy and security risks. Third,  MCP clients, meanwhile, sit at the frontier of ecosystem evolution: some adopt emerging interaction protocols and support multi-server integration, while many remain bound to simple, single-purpose designs. These patterns hint at an ecosystem that is expanding in scale, uneven in quality, and still searching for stable forms of interoperability. Taken together, these concerns highlight the need for a structured investigation of MCP markets, servers, and clients, with a focus on their scale, security, and risk characteristics.

\subsection{Research Questions}

Building on the above motivation, we frame our study around three guiding research questions that capture the scale of the MCP ecosystem, the sensitivity of server-side capabilities, and the security of client-side interactions. These questions aim to provide a comprehensive view of both the opportunities and risks in MCP deployment:

\begin{itemize}
    \item [RQ1] \textbf{(Market): Ecosystem Scale and Growth Potential.} 
What is the current scale of the MCP ecosystem across major markets, and what do observed growth patterns suggest about its future trajectory?

\item [RQ2] \textbf{(MCP Server): Security and Privacy Posture.} 
To what extent is the MCP ecosystem secure and privacy-preserving, and how do structural factors such as dependency choices, maintenance practices, implementation languages, and functional roles collectively shape its overall risk posture?
    \item [RQ3] \textbf{(MCP Client): Client Connection Patterns in MCP Evolution.}
How do the interaction protocols and connection patterns of MCP clients shape the evolutionary trajectory of the ecosystem?

\end{itemize}

%% file: sections/sec04-architecture.tex
\section{Design and Performance Measurement of \mcpc{}}

The objective of this study is to systematically characterize the MCP ecosystem including markets, servers, and clients, which to answer RQ1–RQ3 on ecosystem scale, security/privacy sensitivity, and client interaction risks. To this end, we build a unified measurement pipeline, \mcpc{}, that combines a cross-market crawler with schema normalization, entity deduplication, capability fingerprinting, and connection-mode classification, enabling large-scale indexing and analysis of MCP artifacts. In this section, we first outline the key challenges we encountered and the solutions we adopted to address them (\S\ref{section:resarchquesitons}), before presenting the detailed system design (\S\ref{subsec:design}). We also report on the performance of our measurement tool to demonstrate its scalability and efficiency (\S\ref{subsec:performance}). \looseness=-1

\subsection{Challenges and Insights} 
\label{section:resarchquesitons}

Measuring a decentralized ecosystem is non-trivial: registries differ in data models and access methods (HTML pages, JSON APIs, static catalogs), entries are inconsistently labeled or duplicated across sites, and deployment modes range from locally installed packages to hosted endpoints. We address these challenges with per-registry adapters, schema inference and canonicalization, content hashing for cross-source deduplication, and time-versioned snapshots with rate-limited, robots-aware crawling. This section details the data-collection design, the challenges encountered, and the mitigations that make our measurement reproducible and comprehensive.


\vspace{2mm}
\noindent\textbf{C1. Market-level: Data Heterogeneity and Access Restrictions.} 
MCP servers and MCP clients are indexed by multiple markets (e.g., Glama.ai, MCP.so, PulseMCP). These markets expose project metadata through heterogeneous mechanisms such as JSON APIs, HTML listings, or keyword-based search results. Their metadata schemas also diverge: some markets emphasize quality ratings (e.g., security and license tiers), while others omit such fields entirely. In addition, markets impose strict access constraints, including hard result caps, API rate limits, or Cloudflare-based CAPTCHA challenges. Such heterogeneity and restrictions complicate normalized aggregation, create coverage gaps, and reduce reproducibility of market-level analyses.

\vspace{2mm}
 \begin{mdframed}[backgroundcolor=green!2] 
 {
\noindent\textbf{S1. Modular Adapters and Adaptive Crawling.} 
To handle the heterogeneity and restrictions of different markets, we developed a plugin-based crawler architecture where each market is accessed by a dedicated adapter. This design enables schema normalization while maintaining robustness. Adaptive crawling strategies such as distributing queries across multiple IP addresses, generating keyword variants, and employing semi-automated CAPTCHA handling with cookie reuse, which improve coverage and resilience against access barriers.\looseness=-1
}
\end{mdframed}

\vspace{2mm}
\noindent\textbf{C2. MCP Server-level: Entity Identification and Data Quality.} 
Cross-market entity resolution for MCP servers is hindered by missing or erroneous identifiers. For instance, GitHub URLs—a common linking key—are often absent or contain typos, while supplementary fields such as author names, licenses, or update timestamps are inconsistently recorded. Beyond identification, MCP servers exhibit substantial noise: many entries correspond to placeholder repositories, inactive forks without commits, or projects devoid of executable code. The prevalence of invalid or low-value MCP servers biases statistics and risks overestimating the health of the server ecosystem if not systematically filtered.

\vspace{2mm}
 \begin{mdframed}[backgroundcolor=green!2] 
 {
\noindent\textbf{S2. Multi-Feature Matching and Noise Filtering.} 
Entity resolution for MCP servers cannot rely on single identifiers. Instead, multi-feature matching that combines GitHub URLs, textual similarity of descriptions, author names, and license types is more reliable. Low-confidence cases are escalated for human-in-the-loop verification. In parallel, noise analysis revealed that invalid MCP servers exhibit recurrent patterns, such as placeholder README text, empty directories, or inactive forks. Rule-based filters exploiting these patterns effectively exclude low-value entries, improving dataset quality without discarding valid MCP servers. 
}
\end{mdframed}

\vspace{2mm}
\noindent\textbf{C3. MCP Client-level: Interaction Uncertainty and Evaluation Gaps.} 
Compared with MCP servers, MCP clients are less consistently documented across markets. Their interaction mechanisms with LLMs or MCP servers vary widely: some employ SSE-based communication, others rely on stdio, yet these implementations are rarely described in detail. Furthermore, there is no unified framework for evaluating the adoption, quality, or security posture of MCP clients. This lack of systematic evaluation makes it difficult to understand how MCP clients handle sensitive data flows, whether they risk privacy leakage, and how their architectural choices (e.g., protocol selection) affect robustness and security.

\vspace{2mm}
 \begin{mdframed}[backgroundcolor=green!2] 
 {
\noindent\textbf{S3. Composite Quality Signals and Interaction Profiling.} 
Although no single market provides a complete picture of MCP client adoption or reliability, signals from different markets are complementary. GitHub stars and forks measure community visibility; Glama.ai ratings emphasize security and license compliance; PulseMCP and Smithery report usage statistics. By normalizing and aggregating these signals into composite scores, we obtain a holistic measure of MCP client quality. This composite evaluation also provides a foundation for profiling MCP client–MCP server interaction patterns and assessing potential risks in protocol design and data handling.
}
\end{mdframed}

\subsection{Detailed System Design}
\label{subsec:design}

We developed \mcpc{}, a modular and extensible measurement framework tailored for the heterogeneous MCP ecosystem. \mcpc{} is designed as a pipeline with three main subsystems -- Market Adapter, Server Resolver, and Client Profiler -- linked by a centralized scheduler and backed by a persistent storage layer. A lightweight orchestration service coordinates the scheduling of crawling tasks, retry policies, and data deduplication across these subsystems.

\begin{itemize}
    \item \textbf{Market Adapter.} The first component addresses heterogeneity across MCP marketplaces. It provides a plugin-based adapter layer that unifies diverse data sources (e.g., JSON APIs, HTML pages, or repositories) into a consistent schema. Combined with adaptive crawling strategies, this design ensures broad market coverage while remaining robust against access restrictions such as rate limits or CAPTCHAs.

    \item \textbf{Server Resolver.} The second component focuses on MCP servers, resolving duplicates and filtering invalid entries. By applying multi-feature matching and heuristic noise reduction, it improves data fidelity and enables more accurate assessments of sensitive server functionality, such as authentication or proprietary API exposure.

    \item \textbf{Client Profiler.} The third component targets MCP clients, integrating popularity and quality signals from multiple markets to produce composite evaluations. It further analyzes client interaction patterns (e.g., SSE, stdio), offering a systematic view of connection modes and associated security or privacy risks.
\end{itemize}
 
\begin{wrapfigure}[20]{r}{0.55\columnwidth}
    \centering
    \includegraphics[width=1\linewidth]{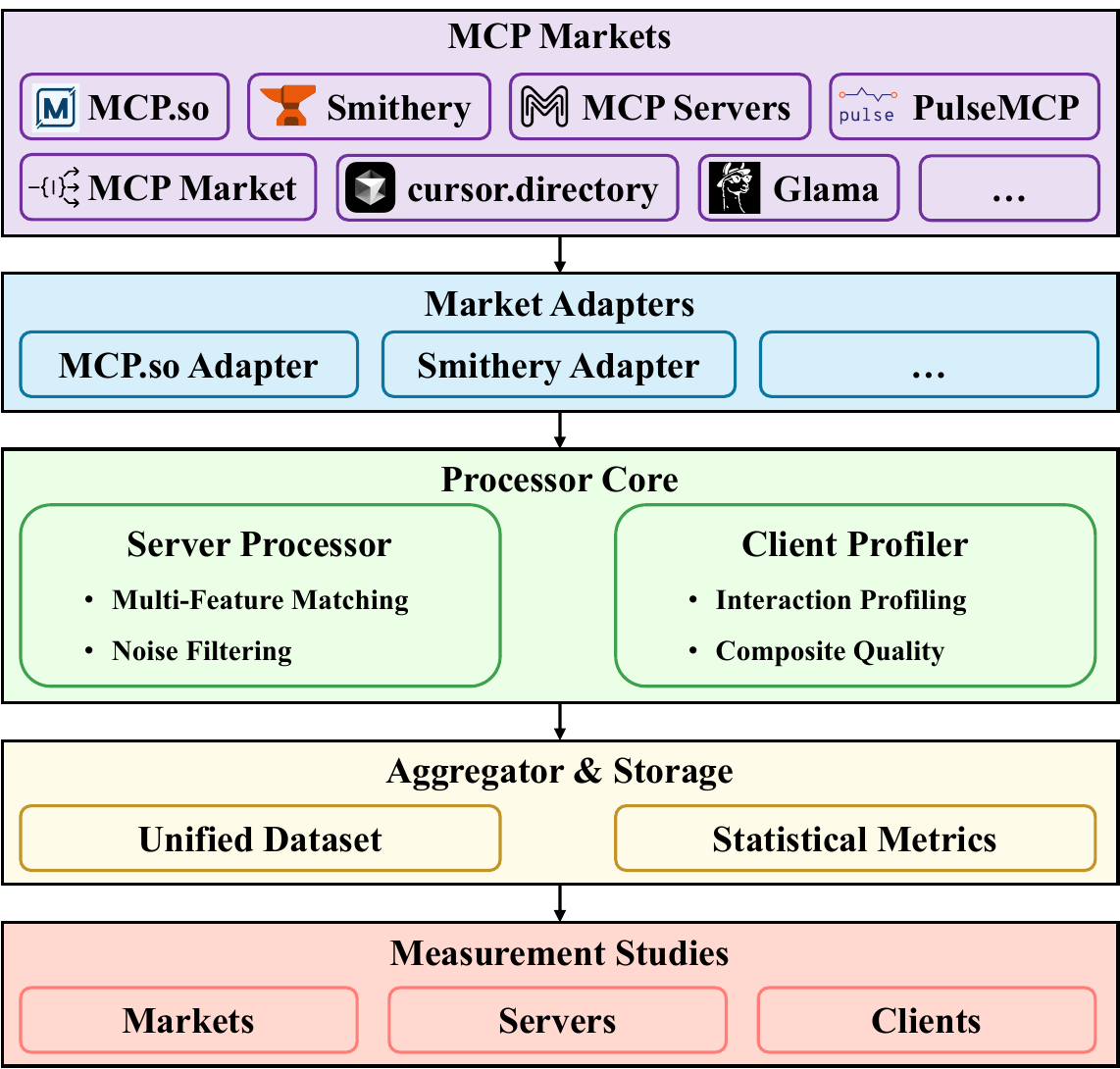}
    \caption{The architecture of \mcpc{}}
    \label{fig:architecture}
\end{wrapfigure}

The overall architecture is depicted in \autoref{fig:architecture}. Each subsystem runs as an independent Python service communicating through a message queue (RabbitMQ in our deployment), allowing horizontal scaling and fault isolation.

\subsubsection{\textbf{Market Adapter}}
To address heterogeneous data formats and access restrictions, the first subsystem is implemented as a plugin-based layer. \autoref{tab:market_adapters} illustrates representative markets, highlighting the heterogeneity that motivates our adapter-based design. Each market (e.g. Glama.ai, MCP.so, PulseMCP) is encapsulated by a dedicated Python module exposing a standard interface: \texttt{fetch()}, \texttt{normalize()}, and \texttt{persist()}. Adapters rely on \texttt{aiohttp} and \texttt{Playwright} for asynchronous HTTP requests and headless browser rendering when encountering dynamic JavaScript pages.

\newpage

A centralized scheduler distributes crawling tasks across a pool of worker nodes. Each task is assigned a market-specific rate limit profile and retry budget. Rotating IP pools and authenticated proxy endpoints mitigate rate limits and CAPTCHAs. If a CAPTCHA challenge is detected, the adapter triggers a semi-automated browser interaction. 
Raw data are normalized into a unified JSON schema with mandatory fields (name, owner, description, URL) and optional fields (ratings, tags, license). 



\subsubsection{\textbf{Server Resolver}}
The second subsystem focuses on MCP servers, with the dual goals of accurate entity resolution and quality assurance. Since MCP server entries are inconsistently documented across markets, \mcpc{} employs a multi-feature matching algorithm that aggregates several signals for entity linking. Specifically, GitHub repository URLs are used as strong identifiers whenever available; textual similarity is computed over project names and descriptions using TF–IDF vectorization with cosine similarity; author and license metadata provide auxiliary features for disambiguating entries with near-identical names; and temporal activity signals (e.g., commit frequency, last update time) capture the likelihood of ongoing maintenance. Each pair of entries receives a composite similarity score, and candidates above a configurable threshold are merged automatically, while borderline cases are flagged for semi-automated, human-in-the-loop verification. This hybrid design ensures high recall without sacrificing precision in deduplication.

\begin{table}[ht]
\centering
\caption{Examples of market heterogeneity and adaptive crawling strategies.}
\label{tab:market_adapters}
\scriptsize
\begin{tabular}{l l l l}
\toprule
\textbf{Market} & \textbf{Data Format} & \textbf{Access Restrictions} & \textbf{\mcpc{} Strategy} \\
\midrule
Glama.ai & JSON API + REST endpoints & Strict API rate limits & Distributed query scheduling with rotating IPs \\
MCP.so & Semi-structured HTML pages & Occasional CAPTCHAs & Session reuse and cached cookies \\
PulseMCP & Static catalogs (web UI) & Limited browsing depth & Recursive crawling with adaptive timeouts \\
\bottomrule
\end{tabular}
\end{table}

\begin{wraptable}{r}{0.6\columnwidth}
\centering
\caption{Representative noise patterns and filtering rules in MCP server entries.}
\label{tab:noise_patterns}
\scriptsize
\begin{tabular}{l l l}
\toprule
\textbf{Noise Type} & \textbf{Indicator} & \textbf{Filtering Rule} \\
\midrule
Placeholder repo    & Only “Init commit”, empty README & Exclude \\
Inactive fork       & Fork with no independent commits & Exclude \\
Abandoned project   & Last commit $>$ 12 months ago    & Exclude \\
Template project    & Matches boilerplate structure    & Exclude \\
Low-content entry   & $<$ 5 source files or missing docs & Flag for review \\
\bottomrule
\end{tabular}
\end{wraptable}

To further improve dataset fidelity, \mcpc{} integrates a noise filtering module derived from our systematic noise analysis. Entries are scored against rule-based heuristics, such as minimum project size (e.g., non-empty README and include 5 source files), repository activity within the last 12 months, exclusion of forks with no independent commits, and removal of placeholder or template projects. \autoref{tab:noise_patterns} summarizes representative noise types and the corresponding filtering rules applied in our system.  Low-confidence entities are filtered out, reducing the prevalence of empty or low-value entries. \looseness=-1

\subsubsection{\textbf{Client Profiler}}
The third subsystem focuses on MCP clients, whose interaction mechanisms and quality signals are fragmented across markets. To provide a unified view, \mcpc{} aggregates complementary indicators into composite quality scores. For open-source clients, GitHub metrics such as stars, forks, issue activity, and release frequency are collected as proxies for adoption and maintenance. Glama.ai entries contribute license compliance and community security ratings, while PulseMCP provides usage statistics that reflect real-world deployment. These signals are normalized and combined through a weighted scoring scheme, allowing clients to be ranked according to adoption breadth, update cadence, and security posture. This integrated view helps distinguish mature, well-maintained clients from experimental or abandoned ones.

In parallel, \mcpc{} implements an interaction profiling module to systematically analyze client–server–LLM connections. Each client is examined for supported communication modes, including SSE, stdio pipes, HTTP streaming, or hybrid protocols. The profiler records handshake sequences, authentication requirements, and session persistence, thereby producing a structured “connection fingerprint” for each client. By comparing these fingerprints across the ecosystem, \mcpc{} reveals dominant patterns (e.g., most hosted clients default to SSE, while local developer tools rely on stdio) as well as edge cases with non-standard or ad-hoc implementations. 
This profiling directly enables risk assessment. For example, persistent SSE sessions without encryption may leak request headers, while stdio-based connections can inadvertently expose sensitive tokens if logging is not sanitized. Clients that mix multiple modes are flagged for closer inspection, since hybrid communication pathways often introduce unexpected data flows.

%

%% file: sections/sec04-performance.tex
\subsection{Performance of \mcpc{}}
\label{subsec:performance}

We evaluate \mcpc{} to understand its efficiency, scalability, robustness, and data quality in practice. Since MCP is an emerging ecosystem, no established baselines exist; thus, our evaluation focuses on absolute metrics that demonstrate \mcpc{}'s ability to support large-scale and reproducible measurements. All experiments were conducted over 14 days across 6 markets (MCP.so, MCP Market, PulseMCP, Smithery, MCP Servers, cursor.directory), resulting in a corpus of 8,060 MCP servers and 341 MCP clients.

\begin{wrapfigure}[11]{r}{0.55\columnwidth}
    \centering
    \includegraphics[width=1\linewidth]{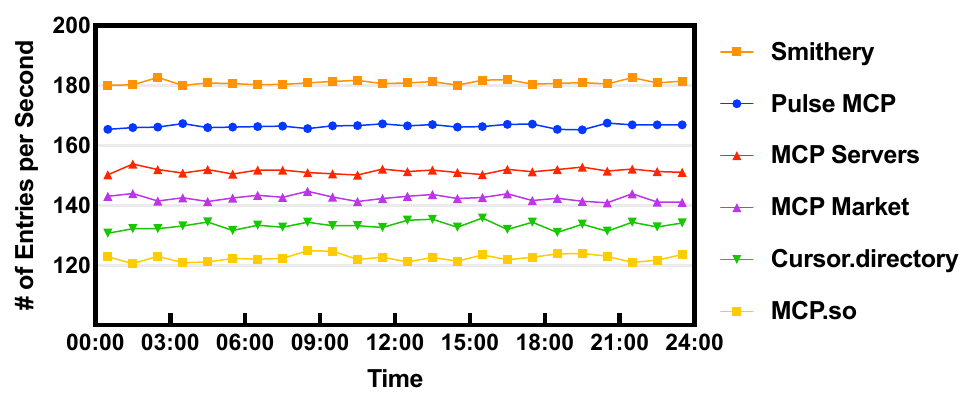}
    \caption{Real-time crawling speed (\# of entries crawled) for a single day.}
    \label{fig:efficiency}
\end{wrapfigure}

\begin{itemize}
\item \textbf{Crawling Efficiency.}
\mcpc{} achieved an average throughput of 147.6 entries/second across all markets. As shown in ~\autoref{fig:efficiency}, the highest performance was observed for Smithery (181.3 entries/second, JSON-based API), while MCP.so were slightly slower (122.1 entries/second, HTML-based) due to parsing overhead. Over the 14-day crawl, \mcpc{} successfully processed 10.2 million entries, showing that the modular adapter design scales effectively to heterogeneous data sources.

\item \textbf{Coverage Across Markets.}
The adaptive crawling design enabled \mcpc{} to index 17,313 distinct entries from 6 markets, including 16,950 MCP server entries and 341 MCP client entries. By distributing queries across IPs and generating keyword variants, the crawler successfully retrieved an additional 18\% entries, compared to a crawling where these optimizations are disabled. Session reuse further allowed \mcpc{} to sustain continuous crawling sessions for up to 36 hours without disruption. These results illustrate that modular adapters and adaptive crawling are essential to maximize ecosystem coverage. 

\item \textbf{Robustness and Stability.}
We measured \mcpc{}’s stability under long-running operations. Over 14 consecutive days, the system maintained an average success rate of 96.7\% per query, with failure rates never exceeding 5\%. Retry mechanisms and session caching proved particularly effective in reducing transient failures, ensuring reproducibility of the crawl. This robustness makes \mcpc{} suitable for continuous monitoring of MCP ecosystem growth.

\item \textbf{Data Quality and Noise Filtering.}
Out of the 16,950 MCP server entries collected, 8,635 were identified as low-value or invalid (e.g., placeholder repositories, inactive forks). Meanwhile, 341 out of 363 MCP client entries collected were identified as valid. This finding suggests that MCP clients tend to be more consistently maintained and of higher quality compared to MCP servers. \mcpc{}’s noise filtering module automatically excluded 50.9\% of these entries. Manual validation of a 500-entry sample confirmed 93.5\% accuracy in invalid server detection. 

\item \textbf{Client Profiling and Composite Evaluation.}
\mcpc{}’s Client Profiler successfully generated composite quality scores for 341 MCP clients. By aggregating visibility metrics (stars, forks), license and security ratings, and usage statistics, \mcpc{} produced stable rankings with variance reduced by 21\% compared to single-source scoring. Interaction profiling revealed that 57\% of clients rely on SSE connections, while 38\% use stdio-based communication. Furthermore, 19\% of clients interact with more than one market-listed MCP server, indicating diverse adoption patterns.
\end{itemize}

%

The performance evaluation of \mcpc{} demonstrates in practice that systematic measurement of the MCP ecosystem is both feasible and informative, yielding insights that extend beyond raw performance metrics. First, the observed efficiency and scalability show that large-scale, multi-market crawling can succeed even in the presence of heterogeneous APIs and restrictive access policies. This indicates that modular adapters and adaptive crawling are not just engineering conveniences, but essential mechanisms for producing datasets that are both comprehensive and reproducible. Second, the large fraction of invalid or low-value MCP servers identified through noise filtering highlights the need for quality safeguards; without them, aggregate statistics would systematically overstate the ecosystem’s vitality. Third, client profiling reveals that interaction modes are already diverging (e.g., SSE vs. stdio), signaling early risks for interoperability and security.

%% file: sections/sec05-market.tex
\section{Measurement of MCP Markets}
\label{sec:markets}

In this section, we conduct a measurement study of the major markets that index and distribute MCP-related projects. Platforms such as Glama.ai, MCP.so, and PulseMCP act as the primary gateways for MCP servers and clients, but they differ widely in scope, metadata practices, and access restrictions. These differences directly shape what projects become visible to the ecosystem and how representative each market truly is\ \cite{extension}. By systematically quantifying market size, examining overlaps across sources, and characterizing the diversity of listed entries, our analysis directly addresses \textbf{RQ1} on the overall scale and distribution of the MCP ecosystem.

\newpage

\subsection{Measurement of Market Growth for Trend Analysis}

\begin{wrapfigure}[12]{r}{0.6\columnwidth}
    \centering
	\includegraphics[width=\linewidth]{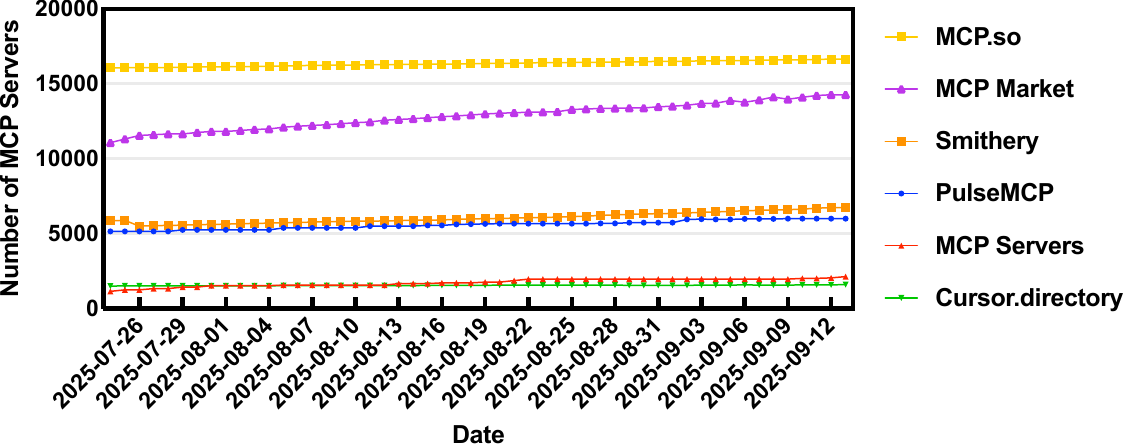}
	\caption{Number of MCP servers in each market.}
    \label{fig:server_number}
\end{wrapfigure}

To understand the dynamics of the MCP ecosystem, it is not sufficient to analyze a static snapshot; instead, a longitudinal view is required to capture how projects appear, evolve, or stagnate across different markets. Market-level adoption and growth patterns provide key signals of ecosystem vitality, concentration, and diversity.
We implemented a 14-day longitudinal crawl using MCP Crawler, targeting six major MCP markets (MCP.so, MCP Market, PulseMCP, Smithery, MCP Servers, cursor.directory). As shown in \autoref{fig:server_number}, for each day between July 26 and September 12, 2025, the crawler queried all listed projects and extracted unique MCP server entries. The resulting time series were aggregated by market and plotted as line charts, where the x-axis represents the date and the y-axis the cumulative number of distinct MCP servers observed. Each line corresponds to one market, enabling comparison of relative market size and growth trends. We filtered out invalid or low-value entries using the rule-based noise removal described earlier. 8,890 entries (52.4\%) out of 16,950 servers were discarded due to patterns such as placeholder repositories, inactive forks, or projects without executable code. The resulting dataset consists of 8,060 valid MCP servers.   Our measurement campaign finally collected 8,401 distinct entries (8,060 MCP servers and 341 MCP clients) in total.

\subsection{Measurement of Ecosystem Scale for Adoption Analysis}

The results in  \autoref{tab:market_stats} reveal that a substantial fraction of MCP projects are invalid or low-value entries. For example, in MCP.so, only 7,223 out of 16,646 server records (43.4\%) were considered valid. The problem is even more pronounced in MCP Market, where just 3,765 out of 14,280 servers (26.4\%) passed validation. PulseMCP shows a similar pattern, with more than 40\% of entries discarded. Overall, across all markets we collected 17,630 raw entries, of which only 8,656 (49.1\%) were valid, meaning that over half of the ecosystem consists of abandoned, placeholder, or otherwise unusable projects.
Specifically,
MCP.so dominated the ecosystem, accounting for 89.1\% of all indexed projects, reflecting its role as the primary hub for servers. Meanwhile, PulseMCP slightly surpasses MCP.so in the number of clients. By contrast, MCP Servers contributed fewer entries but enriched them with detailed usage statistics.
The longitudinal measurement reveals several insights into the structure and dynamics of the MCP ecosystem. While MCP.so remains largely saturated with a stable number of entries, MCP Market shows a steady upward trend, suggesting that it is the primary driver of ecosystem growth. Mid-tier sources such as Smithery and PulseMCP contribute fewer servers but show gradual increases, with PulseMCP further distinguished by providing richer metadata, underscoring a trade-off between coverage and informational depth. Long-tail repositories such as Cursor.directory and MCP Servers contribute relatively small numbers of entries but play a complementary role in improving coverage.

\begin{table} 

	\centering
	\caption{{Market-level statistics of MCP projects collected by \mcpc{}}.}
	\label{tab:market_stats}
    \scriptsize
    \setlength{\tabcolsep}{3.8pt}
	\begin{tabular}{l c c c c c c}
		\hline
		\textbf{Market} & \textbf{Servers (Raw)} & \textbf{Servers (Valid)} & \textbf{Clients (Raw)} & \textbf{Clients (Valid)} & \textbf{Total Entries (Raw)} & \textbf{Total Entries (Valid)} \\
		\hline
		MCP.so & 16,646 & 7,223 & 266 & 266 & 16,912 & 7,489 \\
		PulseMCP & 6,013 & 3,576 & 337 & 279 & 6,350 & 3,855 \\
		MCP Market & 14,280 & 3,765 & 19 & 19 & 14,299 & 3,784 \\
		Smithery & 6,751 & 2,588 & 0 & 0 & 6,751 & 2,588 \\
		Cursor.directory & 1,600 & 1,197 & 0 & 0 & 1,600 & 1,197 \\
		MCP Servers & 2,136 & 997 & 58 & 32 & 2,194 & 1,029 \\
		\hline
		\textbf{Total} & 16,950 & 8,060 & 363 & 341 & 17,313 & 8,401 \\
		\hline
	\end{tabular}
\end{table}


\newpage
\subsection{Measurement of Cross-Market Overlap for Coverage and Diversity Assessment}

\begin{wrapfigure}[16]{r}{0.45\columnwidth}
	\centering
    \includegraphics[width=0.45\textwidth]{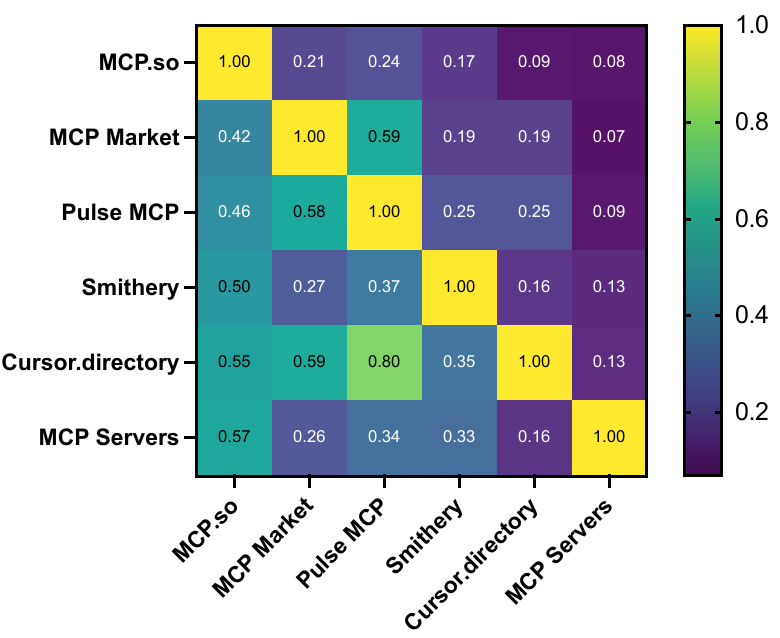}
	\caption{Overlap of MCP projects across markets.}
	\label{fig:market_overlap}
\end{wrapfigure}

A central question in understanding the MCP ecosystem is whether different markets index overlapping sets of projects or instead serve largely distinct communities. To answer this, we performed cross-market entity resolution using a multi-feature matching approach and visualized the results with a pairwise overlap heatmap. Each cell in the heatmap encodes the proportion of shared projects between two markets relative to the size of the market on the row, providing an intuitive view of both bilateral overlap and overall coverage gaps. We chose a heatmap over a raw table because it scales better with multiple markets and makes overlap patterns easier to interpret at a glance. 
The overlap analysis in \autoref{fig:market_overlap} reveals that a large fraction of MCP projects are duplicated across multiple markets, with 32.3\% appearing in more than one platform. However, the duplication is uneven: only 5.5\% of projects are indexed broadly (in four or more markets), while the rest are scattered inconsistently.

\vspace{2mm}
 \begin{mdframed}[backgroundcolor=blue!2] 
 {
\noindent\textbf{Answer to RQ1 (Ecosystem Scale and Growth Potential).} 
Our measurements reveal that the scale of the MCP ecosystem is less substantial than raw market numbers suggest. More than half of indexed projects are abandoned, placeholders, or otherwise low-value, with markets such as MCP Market showing validity rates as low as 26.4\%. Even MCP.so, the dominant hub, appears to have reached saturation, with new growth largely coming from duplication rather than innovation. Cross-market analysis further shows high redundancy but incomplete coverage, which inflates diversity while leaving adoption fragmented. Taken together, these findings suggest that while the ecosystem has reached a visible scale, its growth trajectory remains uncertain and should be interpreted with caution.
}
\end{mdframed}

%% file: sections/sec06-server.tex
\section{Measurement of MCP Servers}
\label{sec:servers}

In this section, we analyze MCP servers, which represent the core service-providing components of the MCP ecosystem. 
To address \textbf{RQ2}, we examine the structural factors that shape the security and privacy posture of the MCP ecosystem from three complementary angles. First, we analyze the library ecosystems on which servers depend, since widely shared dependencies can both propagate vulnerabilities and determine the extent of supply-chain exposure. Second, we study repository characteristics, such as size, code complexity, and maintenance activity, as indicators of whether servers are lightweight and well-maintained or instead oversized and abandoned, thereby influencing patchability and long-term resilience. Third, we measure functionality and implementation choices, focusing on the categories of services and their underlying languages, which directly determine what types of sensitive data may be exposed and how robustly they are handled.




\subsection{Measurement of Library Ecosystems for Supply-Chain Security Assessment}
\label{sec:server:libs}

\begin{figure*}[!htbp]
    \centering
    \includegraphics[width=\textwidth]{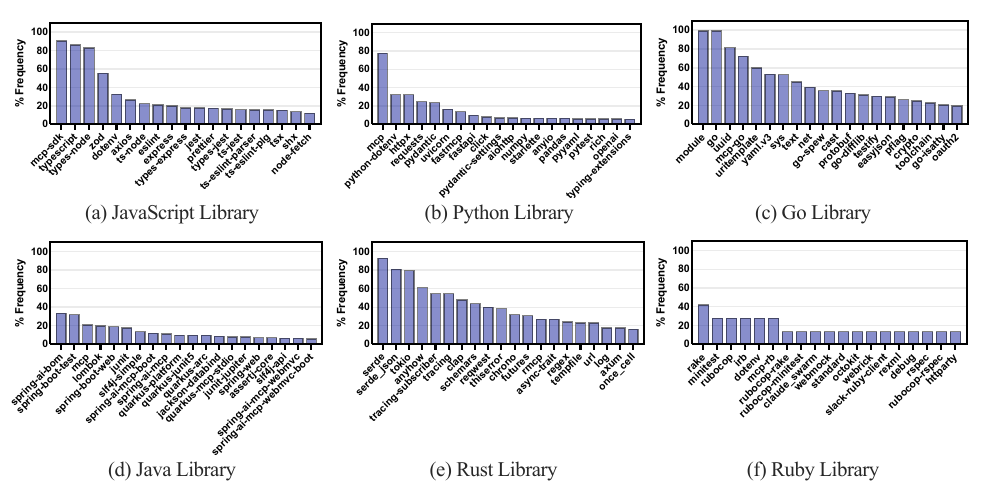}
    \caption{Top 20 libraries used by MCP servers in each language.}
    \label{fig:server_libs}
\end{figure*}

To answer RQ2, we next examine the libraries that MCP servers depend on, with the goal of understanding not only their engineering practices but also their security posture. Libraries are a critical dimension of risk: they may introduce vulnerabilities through unsafe defaults, amplify supply-chain exposure, or conversely, provide safeguards such as schema validation and secure communication\cite{smallworld}. Measuring which libraries are most widely used therefore helps reveal where sensitive functionality is concentrated and whether best practices are being adopted.
\autoref{fig:server_libs} presents the top 20 libraries used by MCP servers in each programming language. To produce this figure, we extracted declared dependencies from server repositories, normalized package names across ecosystems, and computed their frequency. We then visualized the results as bar charts grouped by language, where each bar corresponds to the number of servers importing a given library. This representation makes it straightforward to compare the prevalence of libraries both within and across language stacks.

Our analysis of library usage reveals several security-relevant patterns. 
First, safeguards for input validation are uneven across languages. Python and TypeScript servers frequently rely on schema validation frameworks such as \texttt{pydantic} and \texttt{zod}, which enforce structured input/output and help prevent injection or deserialization flaws. In contrast, Go and Rust servers tend to emphasize serialization through libraries like \texttt{yaml.v3} and \texttt{serde}, but provide fewer explicit validation mechanisms, which may leave developers to implement checks manually and increase the likelihood of unsafe parsing or logic errors. 
Second, the ecosystem shows strong signs of supply-chain monoculture. Java-based servers are overwhelmingly built on the Spring framework (\texttt{spring-boot}, \texttt{spring-core}, \texttt{spring-web}), meaning that a single critical vulnerability such as the 2022 SpringShell remote code execution flaw, could simultaneously impact a large fraction of MCP servers. Similar patterns are visible in Go (\texttt{grpc}) and HTTP client usage (\texttt{axios}, \texttt{requests}), where a bug or misconfiguration in one popular dependency could propagate widely. 
Third, some servers directly integrate with third-party platforms, which expands the attack surface beyond MCP itself. For instance, Ruby projects frequently import connectors such as \texttt{slack-ruby-client} and \texttt{octokit} (GitHub API), which, if misconfigured, could expose API tokens or leak sensitive workspace data. Taken together, these trends indicate that while many MCP servers incorporate good practices like schema validation, the ecosystem remains highly vulnerable to supply-chain attacks and risky external integrations. 

\subsection{Measurement of Repository Characteristics for Maintenance Security}
\label{sec:server:maintenance}
A key security concern for MCP servers lies in their maintenance and complexity, as abandoned or oversized projects may embed unpatched vulnerabilities or expand the attack surface. To evaluate this risk, we analyzed three repository-level features: project size, lines of code,  and commit history,
These three repository-level features provide important signals about the security posture of MCP servers. Project size reflects the storage and dependency footprint: while small projects are easier to audit, very large ones often embed extensive third-party code or datasets, which can increase the likelihood of supply-chain vulnerabilities and complicate patching. Lines of code capture implementation complexity: smaller codebases generally present a narrower attack surface, whereas larger projects raise the probability of hidden bugs or misconfigurations and demand more rigorous review. Finally, commit history serves as a proxy for maintenance activity: actively updated repositories are more likely to apply timely security patches, while projects with sparse or stale commits may rely on outdated libraries and expose users to known vulnerabilities.

\begin{figure}[!htbp]
	\centering
	\begin{minipage}{0.3\textwidth}
		\centering
		\includegraphics[width=1\columnwidth]{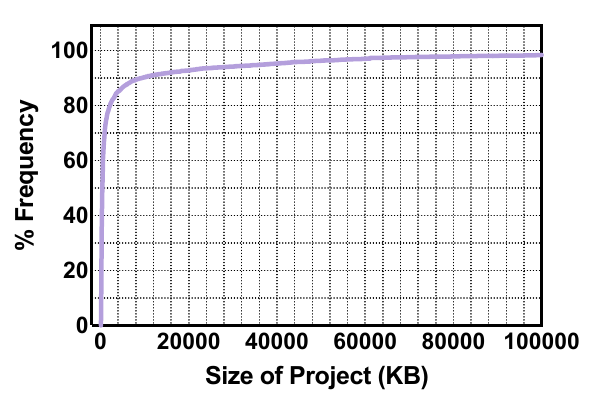}
		\caption{Storage cost distribution of MCP servers.}
		\label{fig:server_size}
	\end{minipage}
	\begin{minipage}{0.3\textwidth}
		\centering
		\includegraphics[width=1\columnwidth]{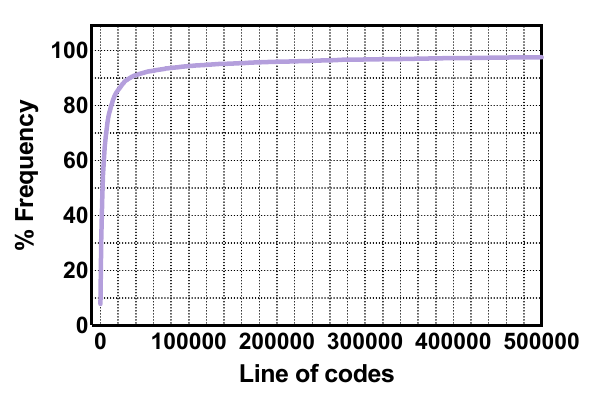}
		\caption{Code lines distribution of MCP servers.} 
		\label{fig:server_loc}
	\end{minipage}
	\begin{minipage}{0.3\textwidth}
		\centering
		\includegraphics[width=1\columnwidth]{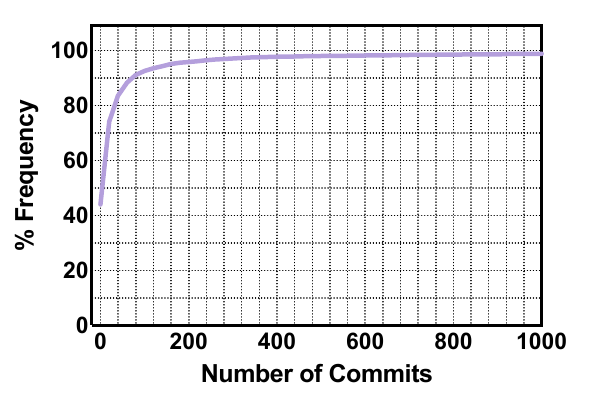}
		\caption{Commits distribution of MCP servers.}
		\label{fig:server_commits}
	\end{minipage}%
\end{figure}

We visualized these metrics through bar and distribution plots (\autoref{fig:server_size}--\autoref{fig:server_commits}), which together provide a multi-faceted view of server activity and engineering practices. The results highlight several insights. First, activity levels are uneven: 40.9\% of servers were updated within the last 90 days and can be considered actively maintained, 37.2\% saw updates in the past year but not the past three months, while 21.9\% have been inactive for more than a year—indicating a significant long tail of abandoned projects that may harbor unpatched flaws. Second, most servers are lightweight (below 50 MB and under 100k LOC), suggesting simple, easy-to-deploy connectors with relatively small attack surfaces. However, a minority of very large projects exhibit heavy data and dependency footprints, raising the likelihood of embedded vulnerabilities. Third, commit distributions confirm that while many servers have limited development activity, a small fraction are intensively maintained and form the ``core'' of the ecosystem, where both innovation and concentrated risk reside.    
Taken together, these findings show that although the MCP ecosystem contains a substantial number of actively maintained projects, the presence of a long tail of outdated or oversized servers poses concrete security risks. Inactive projects are unlikely to receive timely patches and may retain exploitable flaws, while large, dependency-heavy servers expand the attack surface and increase the chance of supply-chain vulnerabilities. As a result, these servers can become attractive targets for attackers, making systematic monitoring of maintenance status a critical requirement for securing MCP deployments.

\subsection{Measurement of  Functionality and Implementation for Exposure Analysis}
\label{sec:server:s_p}

To comprehensively understand the MCP ecosystem from a security perspective, we measure servers along three functional dimensions: categories of exposed services, implementation languages, and API usage. Each of these dimensions provides a distinct lens on potential risks. First, server categories matter because certain functionalities such as authentication, personal data connectors, or proprietary model access, are inherently security-sensitive and expand the attack surface. Second,  the implementation language is closely tied to security posture, since different language vary in their unsafe defaults (e.g., memory safety issues in C/C++ vs. stronger isolation in Rust). Third, sensitive API usage creates a broad attack surface for network, execution, and file-access threats.

\begin{wrapfigure}[11]{r}{0.5\columnwidth}
    \centering
\includegraphics[width=1\linewidth]{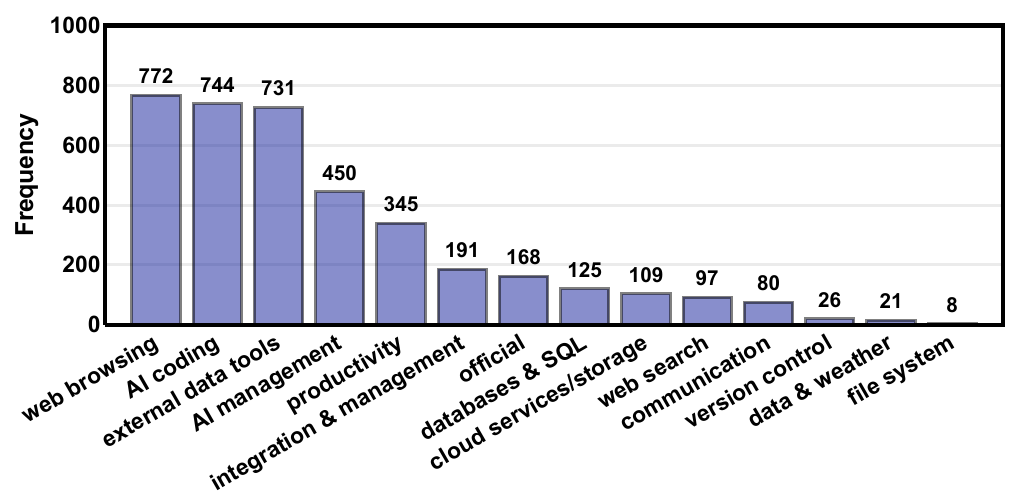}
    \caption{Category distribution of MCP servers.}
    \label{fig:server_category}
\end{wrapfigure}

\begin{wrapfigure}[10]{r}{0.35\columnwidth}
    \centering
\includegraphics[width=1\linewidth]{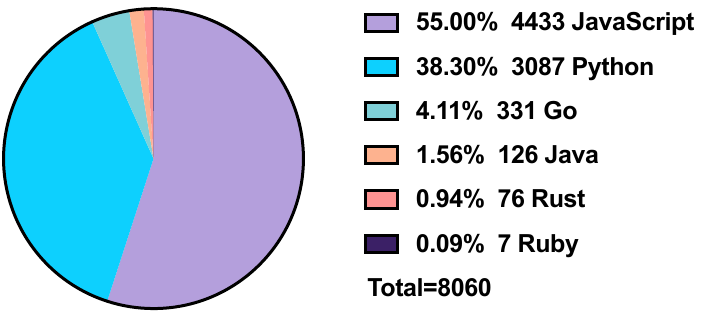}
	\caption{Language distribution of MCP servers.}
	\label{fig:server_lang}
\end{wrapfigure}

To measure functional categories, we collected metadata and documentation from server repositories and classified them based on declared purposes and API endpoints. The resulting taxonomy included productivity tools, integration services, database connectors, authentication systems, and cloud-related services.  \autoref{fig:server_category} visualizes this classification, where each bar corresponds to the fraction of servers belonging to a given functional domain.  For language distribution, we extracted the primary implementation language of each server from repository metadata and plotted the results in  \autoref{fig:server_lang}, which shows the relative prominence of different stacks. 

From a security and privacy perspective, the categories most likely to collect or expose sensitive user data include external data tools (731 servers, 9.1\%), which by design handle structured records and authentication flows, and cloud services/storage (109 servers, 1.4\%), which often involve external APIs and identity connectors that, if misconfigured, could lead to large-scale data leakage. Productivity and collaboration servers (345, 4.3\%) also present privacy risks since they commonly integrate calendars, documents, and communication logs. In addition, web browser content (772, 9.6\%) may capture browsing histories, datasets, or model inputs that contain personally identifiable information. Even smaller categories like communication servers (80, 1.0\%) can expose chat or message data if encryption or access control is weak. The language distribution of MCP servers shows a heavy concentration in JavaScript (55.0\%, 4,433 servers) and Python (38.3\%, 3,087 servers), together accounting for more than 93\% of the ecosystem. This concentration creates a supply-chain monoculture: vulnerabilities in widely used libraries (e.g., npm packages for JavaScript or PyPI modules for Python) could cascade across thousands of servers. Moreover, both languages are highly dynamic, which increases the attack surface for issues such as dependency confusion, prototype pollution, or insecure serialization. In contrast, Go (4.1\%, 331 servers) and Rust (0.9\%, 76 servers) represent smaller fractions but offer stronger safety guarantees, particularly memory safety in Rust and stricter type safety in Go, which may reduce certain classes of vulnerabilities. Java (1.6\%, 126 servers) has a mature security ecosystem but also introduces risks of large-scale exploits when frameworks like Spring or Log4j are affected, as seen in past incidents.

\begin{wrapfigure}[11]{r}{0.6\columnwidth}
    \centering
    \includegraphics[width=1\linewidth]{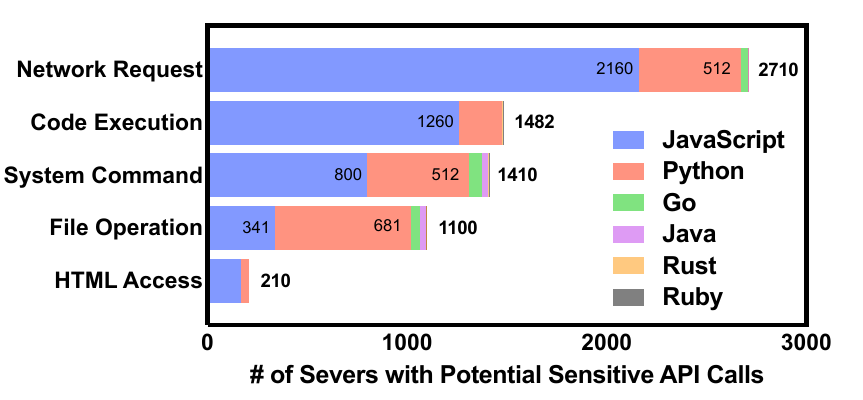}
	\caption{Sensitive API usage of MCP servers.}
	\label{fig:server_api}
\end{wrapfigure}

\begin{table}[!htbp]
    \centering
    \caption{Representative Sensitive APIs in MCP Server Environments.}
    \label{tab:server_api}
    \scriptsize
    \begin{tabular}{p{2cm} p{5cm} >{\raggedright\arraybackslash}p{6cm}}
    \toprule
    \textbf{API Type} & \textbf{Threat Description} & \textbf{Typical API Example} \\
    \midrule
    Network Request & Allows making network connections, potentially enabling data exfiltration, SSRF attacks, or connections to malicious servers & Python: \texttt{requests.get}, \texttt{urllib.request.urlopen}, \texttt{socket.connect} \\
    Code Execution & Permits execution of arbitrary code, allowing attackers to run malicious payloads within the application context & Python: \texttt{eval}, \texttt{exec} \newline JavaScript: \texttt{eval}, \texttt{Function()} \\
    System Command & Executes system commands, possibly leading to unauthorized system control & Python: \texttt{os.system}, \texttt{subprocess.run}, \texttt{subprocess.Popen} \newline JavaScript: \texttt{child\_process.exec}, \texttt{child\_process.spawn} \\
    File Operation & Enables reading or modifying files and directories, which could lead to data leakage or system compromise & Python: \texttt{open}, \texttt{os.remove}, \texttt{shutil.rmtree} \newline JavaScript: \texttt{fs.readFile}, \texttt{fs.writeFile} \\
    HTML Access & Injects untrusted scripts or HTML into responses, potentially compromising client security & JavaScript: \texttt{document.write}, \texttt{element.innerHTML} \\
    \bottomrule
    \end{tabular}
\end{table}

To further assess the security posture of MCP servers, we conducted a static code scan to detect potential invocations of security‐sensitive APIs. The scan used a curated list of library functions commonly associated with high‐risk operations—such as network access, command execution, and dynamic code evaluation—as summarized in \autoref{tab:server_api}. 
\autoref{fig:server_api} summarizes the prevalence of five representative threat categories, with each bar showing the total number of servers containing at least one matched call and internal segments indicating their programming languages. In total, 4,095 servers were flagged as invoking at least one sensitive API. A few cases written in Rust or Ruby were also detected, but their counts are too small to be visible in \autoref{fig:server_api}.
The results show that Network Request APIs are by far the most prevalent, appearing in more than 2,700 servers across languages, dominated by JavaScript (2,160) and Python (512). Code Injection–related calls occur in 1,482 servers, again concentrated in JavaScript (1,260) with smaller but non‐negligible presence in Python (217). Command Execution primitives remain a major concern, with about 1,410 servers referencing OS‐level commands.
Overall, these measurements indicate that MCP servers frequently include code patterns associated with security‐sensitive operations. While such matches do not necessarily imply active exploitation, they reveal a broad potential attack surface that warrants closer auditing. Regular dependency checks, sandboxed execution of network and file operations, and standardized authentication for sensitive API calls are essential to mitigate the risks surfaced by this analysis.

\newpage

\vspace{2mm}
\begin{mdframed}[backgroundcolor=blue!2]
{
\noindent\textbf{Answer to RQ2 (Security and Privacy Posture).}
Our measurements show that the MCP ecosystem is neither uniformly secure nor privacy-preserving, but instead exhibits a mix of encouraging practices and systemic risks. On the positive side, a large fraction of servers remain actively maintained (40.9\% updated within the last 90 days), and many projects incorporate safeguards such as schema validation frameworks (e.g., pydantic, zod) that reduce the likelihood of injection or deserialization flaws. However, a substantial subset of servers include code that invokes \emph{security-sensitive APIs}—for instance, network requests, command execution, or dynamic code evaluation—which, if misused or insufficiently sandboxed, could expose connected environments or external services to compromise. The language distribution amplifies this risk: more than 93\% of MCP servers are implemented in JavaScript or Python, creating a supply-chain monoculture where vulnerabilities in popular npm or PyPI packages could cascade across thousands of deployments.
}
\end{mdframed}

%% file: sections/sec07-client.tex
\section{Measurement of MCP Clients}
\label{sec:clients}
In this section, we conduct a systematic analysis of MCP clients, which function as the crucial interface between end-users, MCP servers, and the underlying LLMs that power model-driven tasks. By design, clients mediate the flow of requests and responses, orchestrating communication and shaping the overall user experience. 
Our measurement focuses on two complementary aspects. First, we examine communication protocols, which uncover whether client–server interactions are converging on a dominant mode or remain fragmented across competing designs. Protocol choice is not only a technical matter but also a signal of standardization, interoperability, and long-term ecosystem trajectory\cite{toolfuzz}. Second, we analyze cross-server usage, capturing whether clients are designed for single-purpose connections or instead support multi-server integration, which enables richer workflows, redundancy, and interoperability across heterogeneous backends. Together, these two perspectives highlight the balance between simplicity and extensibility in client design, and how this balance contributes to the ecosystem’s evolutionary path. 
To support this analysis, \mcpc{} collected a total of 341 valid MCP clients across four major markets. As part of a preprocessing step, we applied strict filtering criteria to remove 22 low-value entries, such as inactive forks, placeholder projects, or duplicates that added noise but no substantive functionality.



\subsection{Measurement of Connection Protocols for Standardization Analysis}

A central aspect of ecosystem evolution lies in how MCP clients establish and sustain connections with servers. Interaction protocols not only determine interoperability but also reflect the broader trajectory of standardization within the ecosystem. If most clients converge on a dominant protocol, this indicates a movement toward a de facto standard that simplifies integration but may reduce diversity. Conversely, the coexistence of multiple protocols suggests an ecosystem still in flux, where different design choices compete for adoption.

\begin{wrapfigure}[10]{r}{0.25\columnwidth}
    \centering
\includegraphics[width=\linewidth]{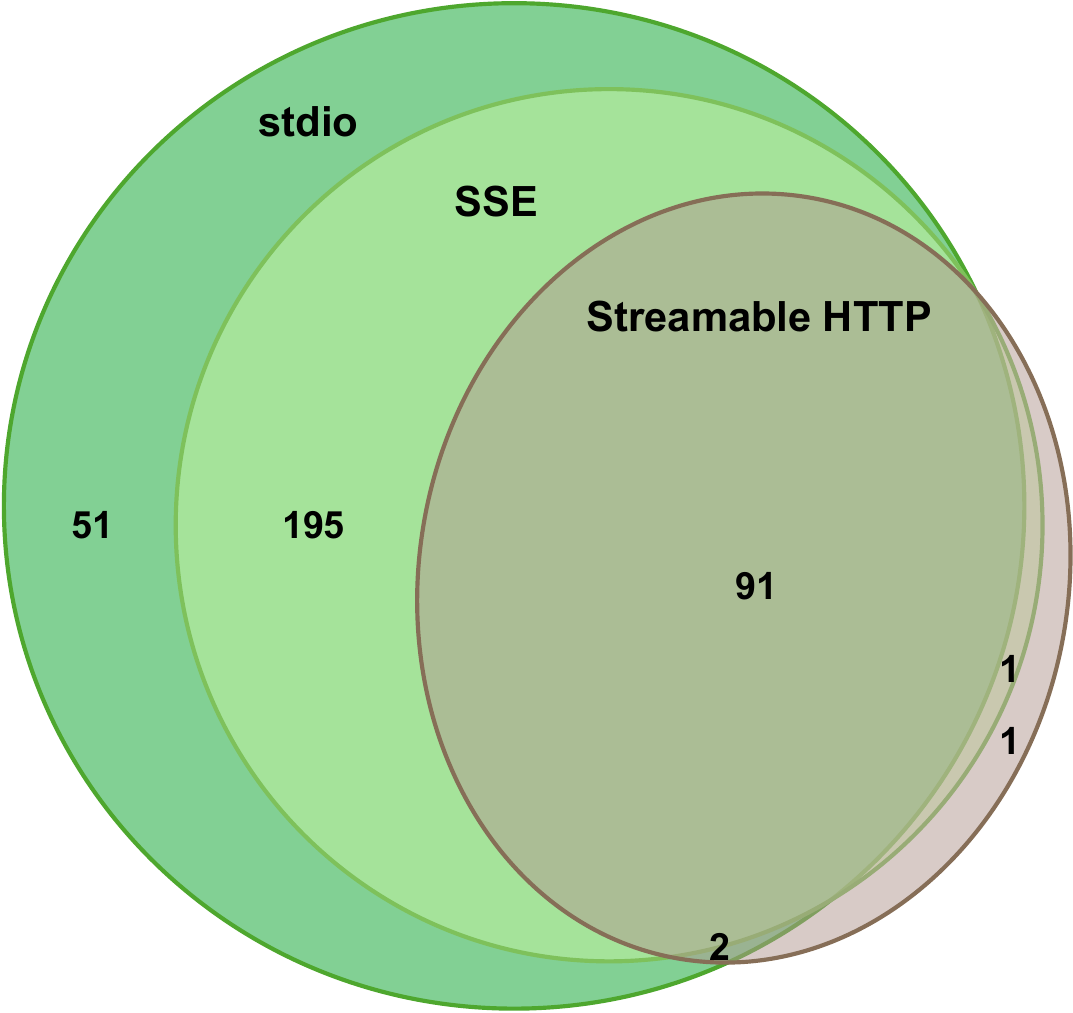}
	\caption{Transport mechanisms of MCP clients.}
\label{fig:client_transport}
\end{wrapfigure}

To characterize client connection patterns, we examined transport support among the 341 valid MCP clients, focusing on the three official communication mechanisms: stdio, Server-Sent Events (SSE), and streamable HTTP. \autoref{fig:client_transport} presents the Venn diagram summarizing their overlap. The results reveal a conservative ecosystem with limited adoption of modern transports. Stdio remains the dominant mechanism, supported by nearly all clients (339 out of 341), reflecting a preference for local, synchronous communication inherited from early MCP prototypes. SSE, although officially deprecated since November 2024 according to the MCP specification\cite{mcp2024transportupdate}, continues to appear widely in co-implementations. In contrast, the officially recommended streamable HTTP transport shows limited adoption: as of August 2025, only 95 of 341 clients support it. This slow transition indicates that the client ecosystem has not kept pace with protocol evolution.
Overall, MCP client development remains lagging behind specification updates. The persistence of legacy stdio and SSE communications, combined with the sluggish uptake of streamable HTTP, reflects substantial engineering inertia within the ecosystem, limiting interoperability and long-term maintainability.


\subsection{Measurement of Client Connection Modes for Usage Dynamics}

\begin{wrapfigure}[10]{r}{0.35\columnwidth}
    \centering
\includegraphics[width=\linewidth]{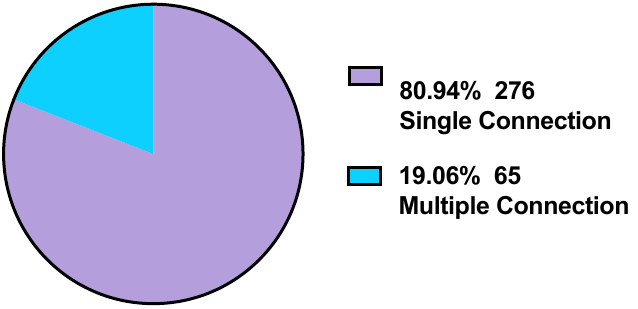}
	\caption{Connection number distribution of MCP clients.}
	\label{fig:client_cross_server}
\end{wrapfigure}
The number of server connections maintained by MCP clients offers a direct view into usage patterns and ecosystem maturity. Many clients are built for single-purpose deployments, where maintaining only one connection simplifies session management, reduces resource overhead, and suffices for vertical use cases such as linking to a single model service or database. By contrast, some clients are designed for integration-oriented scenarios, supporting multiple simultaneous connections to aggregate heterogeneous services, enable redundancy, or provide richer workflows (e.g., in IDEs or enterprise environments). Thus, examining connection numbers helps us understand whether the ecosystem is dominated by simple one-to-one interactions or is moving toward more integrated and versatile client architectures.

We analyzed  MCP clients and extracted metadata on the number of server connections each client supports. Clients were grouped into two categories: single connection and multiple connections. We then calculated the frequency of each group and visualized the results as a pie chart (\autoref{fig:client_cross_server}), where each slice reflects the proportion of clients in that category. The measurement shows that a large majority of clients (80.9\%, 276 clients) maintain only a single connection, while a smaller portion (19.1\%, 65 clients) support multiple connections. This skew suggests that most clients in the current ecosystem are still designed for point-to-point usage, favoring simplicity and ease of deployment. However, the presence of nearly one-fifth of clients with multiple connections reveals a trend toward multi-server integration, which can enable richer functionality, improve resilience, and foster interoperability across heterogeneous services.

\vspace{2mm}
\begin{mdframed}[backgroundcolor=blue!2]
{
\noindent\textbf{Answer to RQ3 (Client Connection Patterns in MCP Evolution).}
Our measurements reveal that MCP client connectivity remains fragmented despite protocol updates. Although SSE was formally deprecated in November 2024, it continues to dominate most implementations, while only 27.9\% of clients have adopted the recommended streamable HTTP transport—evidence of slow migration and technical inertia. On the connection side, 80.9\% of clients link to a single server, with 19.1\% supporting multi-server integration, reflecting early but limited progress toward more complex and interoperable workflows. Overall, client design shows gradual movement toward standardization, yet persistent reliance on legacy mechanisms underscores the ecosystem’s uneven evolution.

}
\end{mdframed}

%% file: sections/sec08-discussion.tex
\section{Discussion}
\label{sec:discussion}

\noindent\textbf{Ethical Considerations.}
Our study was conducted with explicit attention to research ethics. All data we analyzed were drawn from publicly accessible sources, including open repositories, market listings, and metadata exposed through official project listings. At no point did we attempt to bypass authentication mechanisms, crawl private datasets, or interact with servers in ways that could degrade their availability or stability. Importantly, downloading metadata and dependency information from Smithery or other MCP markets does not diminish the commercial or functional value of those platforms; rather, it mirrors the type of visibility already available to any developer or user browsing these resources. 
We did not attempt to probe or exploit server endpoints, ensuring that our methodology posed no threat to operational security. Beyond technical measures, we recognize broader ethical questions around ecosystem measurement, particularly the potential for exposing weaknesses that malicious actors could exploit. To mitigate this, our presentation emphasizes aggregate trends and structural observations, rather than detailed exploit paths.

\vspace{2mm}
\noindent\textbf{Limitations.}
As with any measurement effort, our work has several limitations. First, it represents a snapshot in time, collected over a two-week crawl. The MCP ecosystem is rapidly evolving, and the state of libraries, clients, and servers may shift significantly over weeks or months. Thus, our findings should be seen as characterizing a particular stage of ecosystem growth, not as permanent conclusions. 
Second, our methodology is based primarily on repository metadata and declared dependencies. While these are reliable indicators of engineering practices, they may not fully capture runtime behaviors such as dynamic dependency loading, undocumented features, or proprietary modules. Third, our categorization of server functionality and client protocols necessarily abstracts over complex and heterogeneous implementations. Some projects may straddle multiple categories, and classification errors are possible, though we attempted to minimize them through careful normalization and manual validation.

%% file: sections/sec09-related.tex
\section{Related Work}
\label{sec:related}

The MCP and the broader ecosystem of LLM plugin frameworks are emerging research frontiers that blend protocol design, marketplace dynamics, and security considerations. Because MCP itself was only recently introduced, existing literature remains sparse and fragmented. Previous large-scale measurements of software and application ecosystems\ \cite{measurementgoogleplay, measurementminiapp} have demonstrated the effectiveness of empirical crawling and analysis in revealing structural patterns, maintenance issues, and dependency risks. Inspired by these approaches, we review two closely related lines of work: 
(i) early analyses of the MCP ecosystem, and (ii) open-source agent frameworks and tool libraries that share conceptual similarities with MCP but differ in deployment and measurement scope.

\vspace{2mm}
\noindent\textbf{MCP Ecosystem Analyses.}
The Model Context Protocol (MCP) is a recent open JSON-RPC standard for connecting large language models (LLMs) to external tools and data sources \cite{anthropic2024mcp}. To date, only a handful of academic studies have investigated MCP.
Li et al.\ \cite{li2025urgentlyneedprivilegemanagement} conduct one of the first empirical analyses of real-world MCP-based plugins, revealing that network and system APIs dominate MCP servers and that low-download plugins frequently embed high-risk calls.
Their work highlights the potential of MCP as a unifying protocol for LLM–tool integration but does not report deployment statistics or adoption rates.
Similarly, Anthropic’s official technical documentation \cite{anthropic2024mcp} defines the specification and server lifecycle but provides no usage analytics.
Parallel to MCP, several works have analyzed OpenAI’s plugin-based ecosystems.
Iqbal et al.\ \cite{iqbal2024llmplatformsecurityapplying} propose a systematic attack taxonomy for LLM ``app'' platforms and apply it to OpenAI’s ChatGPT plugin framework, revealing concrete security vulnerabilities.
Ehtesham et al.\ \cite{surveyagentinteroperability} compare MCP with other agent interoperability protocols including ACP, A2A, and ANP, underscoring both its conceptual simplicity and its early-stage standardization challenges.
Prior empirical work on large plugin ecosystems, such as WordPress, has shown that platform–plugin co-evolution strongly influences sustainability and innovation cycles \cite{wordpress}. Similar dynamics are beginning to emerge in MCP markets, where growth and fragmentation occur simultaneously.
These studies focus primarily on protocol description and security, leaving the scale and dynamics of MCP usage largely unexplored.

\vspace{2mm}
\noindent\textbf{Agent Frameworks and Tool Libraries.}
Beyond platform-specific plugins, a variety of open-source frameworks provide standardized mechanisms for tool invocation by LLMs.
LangChain and LlamaIndex define tool schemas for structured API calling \cite{langchain2023,llamaindex2023}.
Hugging Face’s \emph{smolagents} library \cite{wolf2024smolagents} offers a lightweight Python API to build multi-step agentic workflows.
Ge et al.\ conceptualize LLMs as an ``AI Operating System'' capable of hosting multiple agent applications \cite{ge2023aios}.
These frameworks share MCP’s goal of formalizing LLM–tool interactions but their academic treatments remain largely descriptive, with little empirical evidence of real-world adoption.

\vspace{2mm}
 
In summary, existing research underscores the rapid emergence of LLM plugin ecosystems but leaves key questions unanswered. No prior work has quantitatively measured the scale, diversity, or security posture of MCP deployments. Studies of OpenAI plugins and Custom GPTs focus on security but not ecosystem growth, and comparisons across protocol designs remain rare. Our measurement study fills this gap by providing the first large-scale empirical analysis of MCP markets, servers, and clients, while situating these findings alongside comparable LLM plugin frameworks.

%% file: sections/sec10-conclusion.tex
\section{Conclusion}

This paper provides the first comprehensive measurement of the Model Context Protocol ecosystem. By collecting and analyzing 8,401 valid projects across six major markets, we shed light on its scale, security posture, and client interaction patterns. Our findings paint a mixed picture: while MCP has achieved rapid adoption, the ecosystem remains fragile, with over 50\% of projects classified as low-value or abandoned. Servers show promising adoption of input validation frameworks in some languages, but also suffer from supply-chain monocultures, uneven maintenance, and risky exposure of sensitive APIs. On the client side, SSE remains widely used even after its deprecation, and adoption of the newer streamable HTTP transport (27.9\%) has been slow, reflecting the ecosystem’s gradual and uneven protocol evolution.
These insights suggest that MCP is in a transitional stage—widely adopted in appearance but structurally fragile in practice. Moving forward, researchers and practitioners should explore methods for improving ecosystem sustainability, strengthening server security practices, and fostering interoperability among clients. Our dataset and framework aim to support these efforts, offering a foundation for future work on MCP standardization, governance, and security.